\documentclass{aa} 
\usepackage{graphicx}
\usepackage{txfonts}
\usepackage{amssymb,natbib,defs}

\begin{document}
    \title{Obscured and powerful AGN and starburst activities at
           $z\sim$3.5\thanks{This paper makes use of observations collected
           at the European Southern Observatory, Chile, ESO program No. 
           079.A-0522(A), and at the IRAM 30m-Telescope. IRAM is funded by
           the Centre National de la Recherche Scientifique (France), the
           Max-Planck Gesellschaft (Germany), and the Instituto Geografico
           Nacional (Spain). Based on observations obtained with
           MegaPrime/MegaCam, a joint project of CFHT and CEA/DAPNIA, at the
           Canada-France-Hawaii Telescope (CFHT) which is operated by the
           National Research Council (NRC) of Canada, the Institut National
           des Science de l'Univers of the Centre National de la Recherche
           Scientifique (CNRS) of France, and the University of Hawaii. This
           work is based in part on data products produced at TERAPIX and
           the Canadian Astronomy Data Centre as part of the
           Canada-France-Hawaii Telescope Legacy Survey, a collaborative
           project of NRC and CNRS.}}


   \author{M. Polletta\inst{1, 2}, A. Omont\inst{2}
          \and
          S. Berta\inst{3}
          \and
          J. Bergeron\inst{2}
          \and
          C. S. Stalin\inst{4}
          \and
          P. Petitjean\inst{2}
          \and
          M. Giorgetti\inst{5}
          \and
          G. Trinchieri\inst{5}
          \and
          R. Srianand\inst{6}
          \and
          H.~J. McCracken\inst{2}
          \and
          Y. Pei\inst{2}
          \and
          H. Dannerbauer\inst{7}
          }

   \offprints{M. Polletta}

   \institute{INAF-IASF Milano, via E. Bassini, 20133, Italy\\
              \email{polletta@iasf-milano.inaf.it}
              \and
              Universit\'e Paris 6 -- Institut d'Astrophysique de Paris -- CNRS, 98bis blvd. Arago, Paris, 75014 France
              \and
              Max-Planck-Institut f\"ur extraterrestrische Physik, Postfach 1312, 85741 Garching, Germany
              \and
              Indian Institute of Astrophysics, Koramangala, Bangalore 560 034, India
              \and
              INAF - Osservatorio Astronomico di Brera, via Brera 28, 20121 Milano, Italy 
              \and
              IUCAA, Post Bag 4, Ganeshkhind, Pune 411 007, India
              \and
              Max-Planck-Institut f\"ur Astronomie, Königstuhl 17, D-69117 Heidelberg, Germany
              }
   \date{Received 9 June 2008 / Accepted 2 October 2008}

   \abstract
   {}
   {Short phases of coeval powerful starburst and AGN activity during the
    lifetimes of the most massive galaxies are predicted by various models of
    galaxy formation and evolution. In spite of their recurrence and high
    luminosity, such events are rarely observed. Finding such systems,
    understanding their nature, and constraining their number density can
    provide key constraints to galaxy evolutionary models and insights into
    the interplay between starburst and AGN activities.}
    {We report the discovery of two sources at $z$=3.867 and $z$=3.427
    that exhibit both powerful starburst and AGN activities. They benefit
    from multi-wavelength data from radio to X rays from the
    CFHTLS-D1/SWIRE/XMDS surveys. Follow-up optical and near-infrared
    spectroscopy, and millimeter IRAM/MAMBO observations are also available. 
    We performed a multi-wavelength analysis of their spectral energy
    distributions with the aim of understanding the origin of their
    emission and constraining their luminosities. A comparison with other
    composite systems at similar redshifts from the literature is also
    presented.}
   {The AGN and starburst bolometric luminosities are 
    $\sim$10$^{13}$\,\lsun. The AGN emission dominates at X ray, optical,
    mid-infrared wavelengths, and probably also in the radio. The starburst emission
    dominates in the far-infrared. The estimated
    star formation rates range from 500 to 3000\,\msun/yr. The AGN
    near-infrared and X ray emissions are heavily obscured in both sources
    with an estimated dust extinction \av$\geq$4, and Compton-thick gas
    column densities. The two sources are the most obscured and most luminous
    AGNs detected at millimeter wavelengths currently known.}
   {The sources presented in this work are heavily obscured QSOs, but their
    properties are not fully explained by the standard AGN unification model. 
    In one source, the ultraviolet and optical spectra suggest the presence
    of outflowing gas and shocks, and both sources show emission from hot
    dust, most likely in the vicinity of the nucleus. Evidence of moderate,
    AGN-driven radio activity is also found in both sources. Based on the
    estimated stellar and black hole masses, the two sources lie on the
    local $M_{BH}-M_{bulge}$ relation. To remain on this relation as they
    evolve, their star formation rate has to decrease or stop. Our results
    support evolutionary models that invoke radio feedback such as the star
    formation quenching mechanism, and suggest that such a mechanism might
    play a major role also in powerful AGNs.}

   \authorrunning{Polletta et al.}
   \titlerunning{Obscured and powerful AGN and starburst at $z$$\sim$3.5}
   \maketitle
%

\section{Introduction}\label{intro}

The similarity between the star formation history and the space density of
active galactic nuclei (AGNs) over cosmic
time~\citep[e.g][]{hartwick04,hasinger05} and the correlation in nearby
galaxies between black hole (BH) mass and bulge
mass~\citep{ferrarese02,gebhardt00} indicate that star formation in a galaxy
is related to the growth of its BH~\citep[but see ][]{shields08}.
Theoretical
models~\citep{granato01,granato04,dimatteo05,hopkins05b,springel05a} explain
the link between star formation and AGN activity. They show that gas-rich
galaxy mergers are viable precursors to the formation of both massive
galaxies and super massive BHs (SMBHs). According to these models, QSOs and
(sub)-millimeter galaxies (hereinafter SMGs), the most intense sites of star
formation at high-$z$, represent different stages in an evolutionary
sequence, and the obscured growth phase of the BH coincides with the
transition from the SMG to the QSO stages~\citep{page04,stevens05}. QSOs are
thus expected to be the end-product of the rapid growth of SMBHs seen soon
after a phase of high star formation, as seen in
SMGs~\citep{sanders88,granato01}. Although these models claim to be
successful in reproducing and explaining several
observations~\citep{hopkins05c,hopkins06a,li07,chakrabarti07}, more
observations are necessary to test their predictions, constrain their
parameters, and provide a physical base to some of their
assumptions~\citep[see e.g.][]{marulli08}.

Previous studies of SMG populations and high-$z$ starbursts have revealed
the presence of obscured AGN activity in $\sim$30--40\% of these
systems~\citep[e.g.][]{page04,chapman05,stevens05,alexander05a,yan07,pope08},
giving support to a link between the two activities. However, it is not
clear how star formation and BH growth are linked, what their relative
timescales are, and whether they influence each other directly through a
feedback mechanism or indirectly, e.g. by consuming the available cold gas.
The number of well studied composite systems is still low, and it is often
difficult to separate and quantify the contribution from different energy
sources~\citep[see e.g. the variety of models proposed to explain the
properties of F\,10214+4724;][]{rowan-robinson93,teplitz06,efstathiou06}. In
order to quantify their contribution to the bolometric luminosity, the
multi-wavelength spectral energy distributions (SEDs) of such systems need
to be measured and modeled. Because of the difficulty of obtaining a full
multi-wavelength coverage, AGN, starburst galaxies, and composite systems
with well sampled SEDs and known spectra are often used to derive
correlation between bolometric luminosities and measurements at specific
wavelengths or of spectral features~\citep[e.g.][]{sajina08,polletta08a}. In
case of powerful starburst galaxies, like SMGs, the AGN is often highly
obscured and requires either ultra deep X-ray
observations~\citep{alexander05b}, spectropolarimetric
observations~\citep{goodrich96}, or mid-infrared (MIR) spectroscopic
observations~\citep{pope08} to be revealed.

In order to investigate the link between star formation and AGN activity, it
is thus important to study systems where both activities are taking
place, and collect multi-wavelength measurements to constrain the starburst,
i.e. the star formation rate (SFR), and AGN, i.e. accretion rate,
luminosities, as well as their Eddington ratio, and stellar and gas masses.

Here, we investigate the properties of two rare millimeter (mm) bright
obscured QSOs at high-$z$ discovered in a wide multi-wavelength survey.
Spectroscopic data as well as the full SED, from X ray to radio wavelengths
are available for both objects and are analyzed to constrain the origin of
their luminosity and investigate their nature. Throughout this paper, we
adopt a flat cosmology with H$_0$ = 71 \kmsMpc, $\Omega_{M}$=0.27 and
$\Omega_{\Lambda}$=0.73~\citep{spergel03}.

\section{Target selection and observations}\label{obs}

In this work, we analyze the properties of two sources that were selected as
$g$ drop-out sources, thus $z$$\simeq$4 candidate, in the 0.9\,deg$^2$
CFHTLS survey D1\footnote{\url{http://www.cfht.hawaii.edu/Science/CFHTLS/}}
(RA = 02\hr 26\min, Dec = $-$04\deg 30\arcmin) and as very bright MIR
sources in the XMM-LSS field of the {\it Spitzer} Wide-area InfraRed
Extragalactic Legacy
survey~\citep[SWIRE\footnote{\url{http://swire.ipac.caltech.edu/swire/swire.html}};][]{lonsdale03}.
The sources IAU official names are \object{SWIRE2 J022550.67$-$042142.2}
(SW022550 hereinafter), and \object{SWIRE2 J022513.90$-$043419.9} (SW022513
hereinafter). We applied a modified version of the drop-out selection
technique developed by~\citet{steidel99} for $z>$3--4 Lyman-break sources.
In addition to the optical $ugr$ colors and the source extension, this
selection technique takes into account the 3.6\,$\mu$m--4.5\,$\mu$m
color~\citep[Bergeron et al., in prep.; see also][]{siana08}. No other
sources with these properties were found in the 0.9\,deg$^2$ CFHTLS-D1/SWIRE
field. Because of their large 24$\mu$m fluxes (F$_{24\mu m}>$1\,mJy),
exceptionally red infrared (IR) SEDs ($\alpha_{IR}<$$-$2.5 where
F$_{\nu}\propto \nu^{\alpha_{IR}}$ over the observed wavelength range
3--24\,$\mu$m), and high-$z$, we selected both sources for observations at
1.2\,mm with the Max Planck Millimeter Bolometer (MAMBO)
array~\citep{kreysa98} at the Institut de Radioastronomie Millim\'etrique
(IRAM) 30m telescope. Follow up spectroscopic observations were carried out
and broad-band photometric data from X ray to radio wavelengths were already
available in the field.

\subsection{Optical and infrared imaging}

Broad-band photometric data at optical and IR wavelengths are
available for both sources from various surveys. Optical data
in 5 broad-bands, $ugriz$, were provided by the CFHTLS survey D1 (data release
T0004\footnote{\url{http://terapix.iap.fr/rubrique.php?id\_rubrique=241}}).
Near-IR (NIR) data in the J and K bands were obtained from the UKIRT Infrared Deep
Sky Survey~\citep[UKIDSS\footnote{\url{http://www.ukidss.org/}} data release
3;][]{dye06,lawrence07}. Infrared data in all seven \spitzer\ bands, from
IRAC~\citep[3.6, 4.5, 5.8, and 8.0\,$\mu$m; ][]{fazio04} and MIPS~\citep[24,
70, and 160\,$\mu$m; ][]{rieke04}, are available from the SWIRE survey. The
SWIRE data correspond to the latest internal catalogs that will be released
as part of the Data Release 5~\citep[DR5; for details on
the data reduction see][]{surace05}. The total measured magnitudes and
fluxes in each band are reported in Table~\ref{basic_data}.

\subsection{MAMBO observations}

Observations at the IRAM 30m telescope were carried out during March--April
and November 2006, using the 117 element version of the MAMBO array
operating at a wavelength of 1.2\,mm (250 GHz). We used the standard on-off
photometry observing mode, chopping between the target and sky at 2\,Hz, and
nodding the telescope every 10 or 20\,s. On-off observations
were typically obtained in blocks of 6 scans of 16 or 20 10s-subscans each,
and repeated in later observing nights. The atmospheric transmission was
intermediate with $\tau$(1.2\,mm)=0.1--0.4. The absolute flux calibration
was established by observations of Mars and Uranus, resulting in a flux
calibration uncertainty of about 20\%.

On average, the noise of the channel used for point-source observations was
about 35--40 mJy/$\sqrt{t}$/beam, where $t$ is the exposure time in seconds,
consistent with the MAMBO time estimator for winter conditions. The two
sources were observed as part of a larger program targeting more than 100
starburst, AGN, and composite sources selected in the SWIRE
fields~\citep[][Polletta et al., in prep., Fiolet et al., in
prep.]{lonsdale08}. The resulting 1.2\,mm fluxes are 4.7$\pm$0.8\,mJy and
5.5$\pm$0.7\,mJy for SW022550, and SW022513, respectively.

\subsection{Optical spectroscopy of SW022550}

An optical spectroscopic observation of SW022550 was carried out with the
AAOmega system~\citep{sharp06} on the 3.9-m Anglo-Australian Telescope, on
September 27, 2006, as part of a multi-object spectroscopic program of QSOs
in the CFHTLS (Petitjean et al., in prep., Yu et al. in prep.). The 580V and
385R grating were used in the blue and red arms of the spectrograph,
respectively, yielding R $\sim$ 1300. The total integration time was 3\,h,
divided into six 30\,min exposures.

Data reduction was performed using the AAOmega data reduction pipeline
software DRCONTROL. The two dimensional images were flatfielded, and the
spectrum was extracted (using a gaussian profile extraction), wavelength
calibrated and combined within DRCONTROL. The final spectrum was flux
calibrated using the broad-band photometric measurements in the $g$ and $r$
bands after deconvolving the spectrum with the filter bandpasses. This
calibration assumes that the source optical emission has not significantly
varied between the spectroscopic and photometric observations.

\begin{figure}
   \centering
   \includegraphics[width=9cm]{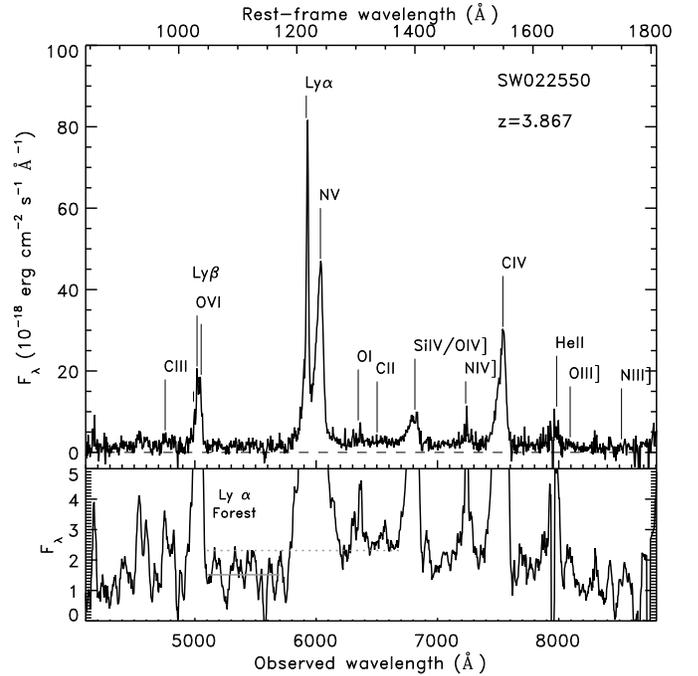}
      \caption{AAOmega optical spectrum of SW022550. The expected emission
       lines are labeled. The horizontal dashed line corresponds to a null
       flux. The bottom panel shows the same spectrum but smoothed with a
       kernel of 7 pixels to better illustrate the continuum emission. The top
       dotted grey line represents the mean continuum level at rest-frame
       $\lambda$=1275--1380\,\AA, and the bottom solid grey curve the mean
       continuum level at $\lambda$=1050--1170\,\AA. }
         \label{spe_xmm55}
   \end{figure}

The optical spectrum of SW022550 shows several strong emission lines, e.g.
\lya, \nv, \civ, the \ovi\ doublet (see Figure~\ref{spe_xmm55}). The
continuum emission is weak ($<$2$\sigma$), and a drop in the continuum
emission is visible at rest-frame wavelengths below 1216\AA\ due to
intergalactic medium (IGM) absorption. The bottom panel of
Figure~\ref{spe_xmm55} shows the decrease in the continuum emission at
rest-frame wavelengths $\lambda$$<$1216\,\AA. Absorption features due to the
Lyman\,$\alpha$ forest are also visible. Since the line spread function
(LSF) derived from the lamp spectra is consistent with a Gaussian, we used
such a profile to fit the emission lines. From the fits we measured the
total line flux, the central wavelength, the FWHM in \AA, and the flux of
the local linear continuum. The FWHM in \kms\ and the rest-frame equivalent
width ($W_{\lambda}$) in \AA\ are then derived from these parameters after
correcting for instrumental resolution. The instrumental FWHM is 3\,\AA\ at
$\lambda$$<$5500\,\AA, and 5\,\AA\ at $\lambda$$>$5500\,\AA. The line flux,
central wavelength, corresponding redshift, rest-frame FWHM and rest-frame
$W_{\lambda}$ of the main visible lines are reported in Table~\ref{spe_tab}.
Uncertainties are determined using a Montecarlo method and assuming a
Gaussian noise given by the r.m.s. associated with the continuum where the
fit is performed. A broad component, in addition to a narrow one, is
measured in correspondence of the \lya, \nv, \siiv, and \civ\ lines. All these
lines show an asymmetric component, similar to a blue wing. To fit such an
asymmetric profile, in some cases, e.g. \lya, and \nv, we fit multiple
Gaussian components. Based on the central wavelength of the optical lines
listed in Table~\ref{spe_tab}, we estimate a redshifts of 3.867$\pm$0.009.

\begin{figure}
   \centering
   \includegraphics[width=9cm]{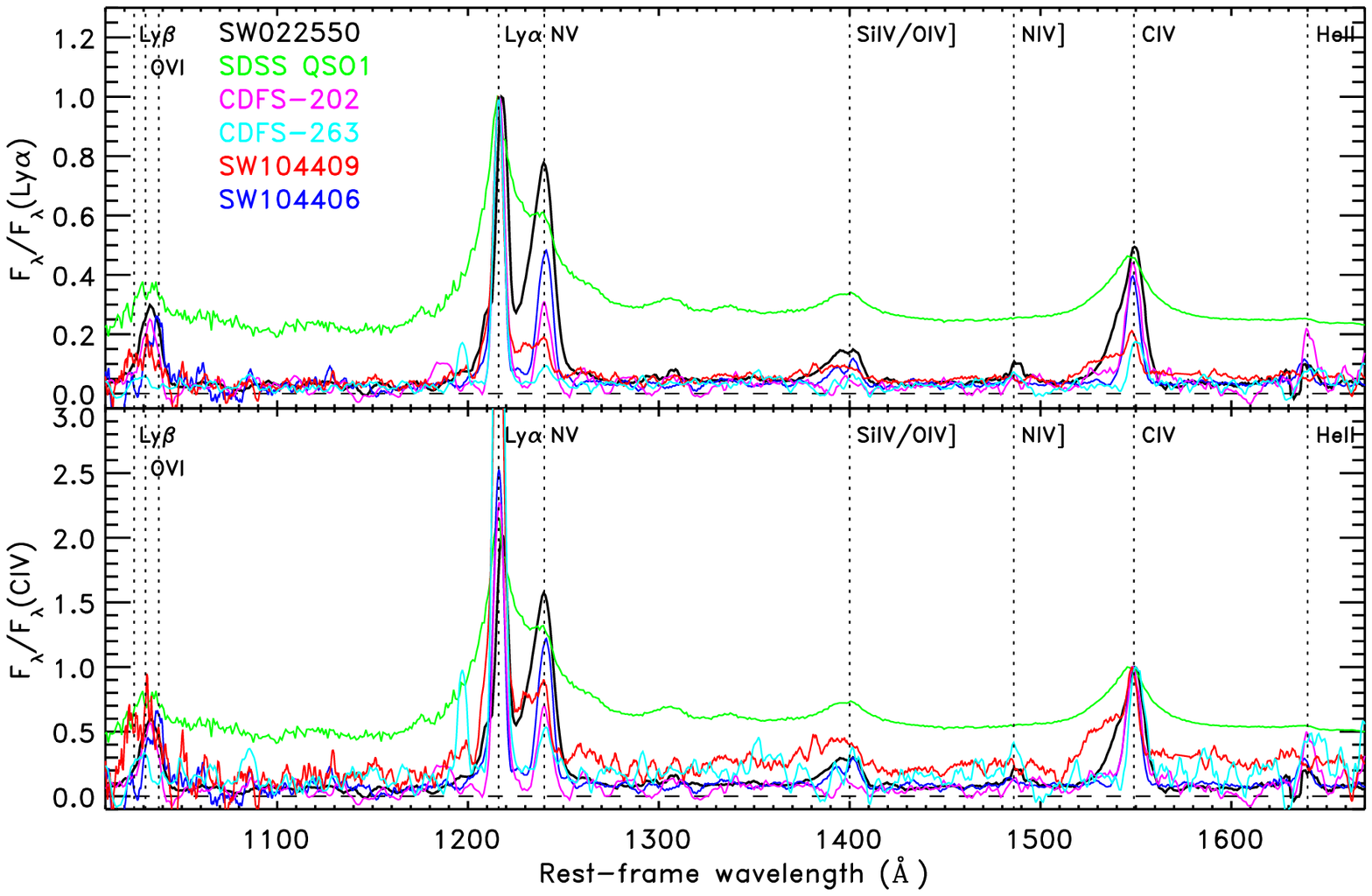}
      \caption{{\it Top panel: } Optical spectra normalized at the flux at
       1216\,\AA\ in the rest-frame of SW022550 (black thick line), of the
       type 2 QSOs \object{CDFS-263} (cyan line) and \object{CDFS-202}
       (magenta line) from~\citet{szokoly04}, and \object{SW104406} (blue
       line), and \object{SW104409} (red line) from~\citet{polletta06}, and
       medium composite spectrum of SDSS type 1 QSOs~\citep[green
       line;][]{vandenberk01}. {\it Bottom panel:} As top panel, but
       normalized at the flux at 1549\,\AA\ rest-frame.  The main emission
       lines are labeled and marked with dotted vertical lines. The
       horizontal dashed line corresponds to a null flux. All spectra have
       been smoothed for clarity.}
         \label{spe_comparison}
   \end{figure}

The emission lines are dominated by their narrow (FWHM$\sim$1000\,\kms)
components with FWHMs typical of type 2 QSOs~\citep[$\leq$1500\,\kms; see
e.g.][]{baldwin88,norman02,mainieri05}. Broad components are also present
with FWHM$\sim$2000--9000\,\kms as observed in type 1
QSOs~\citep[$\geq$2000\,\kms; see e.g.][]{richards04}. The equivalent widths
of the emission lines in the optical spectrum of SW022550 are larger than
observed in type 1 QSOs, and more similar to those observed in other type 2
QSOs~\citep[see e.g. SMM\,02399$-$0136, CDFS-263, or CXO\,52;
][]{ivison98,mainieri05,stern02}. In Figure~\ref{spe_comparison}, we compare
the optical spectrum of SW022550 with a composite spectrum of a large sample
of type 1 QSOs and with the spectra of four type 2 QSOs at high-$z$ from the
literature. We show the spectra normalized at the \lya\ peak on the top
panel, and at the \civ\ peak in the bottom panel. The composite type 1 QSO
spectrum corresponds to the median composite spectrum of 2204 QSO spectra
from SDSS~\citep{vandenberk01}. The type 2 QSOs have been selected because
of the availability of their optical spectra from the literature and
high-$z$. They are \object{CDFS-202}, and \object{CDFS-263} from
CDFS~\citep{szokoly04}, and \object{SW104406} and \object{SW104409} from the
SWIRE/\chandra\ survey~\citep{polletta06}. This comparison shows the wide
range of line ratios that can be observed in type 2 QSOs. The main
differences reside in the strength of the \nv, and \civ\ emission lines, and
the blue wings observed in the main emission lines. We will discuss below
two possible explanations for the strength of the \nv, and \civ\ lines, i.e.
high metalicities and shocks. Interestingly, there is another object in the
type 2 QSO sample from the literature with similar asymmetric lines, i.e.
SW104409. This source is also characterized by a similar optical-IR SED to
SW022550, with an unusual optical blue continuum with strong AGN emission
lines, and red optical-MIR colors. A possible explanation for the presence
of blue wings is extinction due to dust mixed with infalling or outflowing
ionized gas~\citep{osterbrock89}. Since the SED in both cases indicate that
the AGN is obscured, it is also plausible that the blue continuum and the
broad blue wings are due to scattered light~\citep[see e.g.][]{zakamska06}.

SW022550 is characterized by large \nv/\civ\ and \nv/\heii\ flux ratios.
These flux ratios are often used to estimate the gas metalicity in QSOs,
from either the broad or the narrow line region~\citep[BLR and NLR; see
][and references therein]{nagao06}. High values usually imply high
metalicities~\citep[see e.g.][]{hamann93,vernet01a,norman02}. Based on the
photoionization model in~\citet{hamann93}, we derive a metalicity
$Z$=4$Z_{\odot}$. However, not all the emission line ratios of the spectrum
of SW022550 are explained by NLR photoionization models~\citep[see
e.g.][]{nagao06,groves04}. A possible cause of this discrepancy is the
presence of broad components. It is also possible that the emission lines
are produced by shocks in addition or rather than by photoionization. This
hypothesis is supported by the strength of some lines, like \civ, \ovi, and
\nv. For example, \civ\ is predominantly produced in the cooling zone of
shocks and it is expected to be stronger in shocks than in photoionization
processes. Indeed, the \civ/\heii, \nv/\heii, and \nv/\civ\ flux ratios like
those observed in SW022550, are higher than predicted by photoionization,
and can be explained by shocks~\citep{allen08}, or by the coexistence of
shocks and AGN photoionization, without requiring largely super-solar
metalicities~\citep{moy02}. Also the presence of \ovi\ can be considered as
evidence of shocks, although other explanations are also possible. 
To confirm whether SW022550 is characterized by high metalicity and better
constrain its value, or whether shocks are in part responsible for the
observed emission lines, it would be necessary to measure other emission
lines, especially at rest-frame optical wavelengths.

In summary, the optical spectrum of SW022550 shows spectral features that
are also observed in other obscured (type 2) AGNs at high-$z$, but are not
explained by the standard AGN unification model~\citep{antonucci93}. In
particular, a high metalicity or the presence of shocks and a significant
scattered fraction at rest-frame UV wavelengths are necessary to fully
explain the observed spectrum. Note that these hypothesis can not been
fully tested with the available data and other scenarios are not ruled out.

\begin{table} 
\begin{minipage}[t]{\columnwidth}
\caption{Multi-wavelength data of SW022550 and SW022513} \label{basic_data}
\centering 
\renewcommand{\footnoterule}{} 
\begin{tabular}{llrr} 
mag or Flux\footnote{\small X ray fluxes are in 10$^{-15}$\,\ergcm2s.
Optical magnitudes are total magnitudes in the AB systems in the $ugriz$
filters from the CFHTLS survey D1. UKIDSS J and K fluxes, and IRAC
(3.6--8.9\,$\mu$m) fluxes are in $\mu$Jy. MIPS (24--160\,$\mu$m), MAMBO
(1.2\,mm) and radio (20\,cm) fluxes are in mJy. X-ray upper limits
correspond to 3$\sigma$, optical upper limits correspond to 80\%
completeness, IR upper limits correspond to 5$\sigma$. The radio data are
from the VIMOS VLA survey~\citep{bondi03,bondi07}.}
                 & Project/Telescope&      SW022550        &               SW022513  \\
\hline\hline     
F$_{0.5-2\,keV}$ & XMDS/XMM         & $<$2.5               & 1.1$\pm$0.8 \\
F$_{2-10\,keV}$  & XMDS/XMM         & $<$20                & 7.8$\pm$4.1 \\
m$_\mathrm{u}$   & CFHTLS/CFHT      &     25.211$\pm$0.077 &                $>$26.00 \\               
m$_{g^{\prime}}$ & CFHTLS/CFHT      &     22.944$\pm$0.007 &        24.432$\pm$0.033 \\
m$_{r^{\prime}}$ & CFHTLS/CFHT      &     21.648$\pm$0.003 &        23.287$\pm$0.014 \\
m$_{i^{\prime}}$ & CFHTLS/CFHT      &     22.949$\pm$0.004 &        22.851$\pm$0.012 \\
m$_z$            & CFHTLS/CFHT      &     22.408$\pm$0.021 &        22.503$\pm$0.027 \\               
F$_J$            & UKIDSS/UKIRT     &       3.6$\pm$0.4    &          3.1$\pm$0.6  \\
F$_K$            & UKIDSS/UKIRT     &       7.7$\pm$0.6    &         12.7$\pm$0.6   \\
F$_{3.6\mu m}$   & SWIRE/\spitzer\  &             15$\pm$1 &                16$\pm$1 \\
F$_{4.5\mu m}$   & SWIRE/\spitzer\  &             23$\pm$1 &                20$\pm$1 \\
F$_{5.8\mu m}$   & SWIRE/\spitzer\  &                $<$58 &                28$\pm$4 \\
F$_{8.0\mu m}$   & SWIRE/\spitzer\  &            239$\pm$7 &               180$\pm$5 \\
F$_{24\mu m}$    & SWIRE/\spitzer\  &        3.32$\pm$0.02 &           2.35$\pm$0.02 \\
F$_{70\mu m}$    & SWIRE/\spitzer\  &                $<$24 &                   $<$24 \\
F$_{160\mu m}$   & SWIRE/\spitzer\  &               $<$126 &                  $<$126 \\
F$_{1.2mm}$      & IRAM/MAMBO       &        4.70$\pm$0.77 &           5.53$\pm$0.72 \\
F$_{20cm}$       & VIMOS/VLA        &        0.14$\pm$0.02 &           0.35$\pm$0.02 \\
F$_{50cm}$       & VIMOS/GMRT       &        0.41$\pm$0.04 &           0.62$\pm$0.04 \\
\hline
\end{tabular}
\end{minipage}
\end{table}

\subsection{Near-infrared spectroscopy}

Both targets were observed with the ISAAC instrument~\citep{moorwood98} on
Antu (VLT-UT1) in low-resolution (LR) mode, using the 1024$\times$1024
Hawaii Rockwell array of the Short Wavelength arm. 
Observations were carried out with a 1\arcsec\ slit with two grating blaze
angles, approximately covering the SH ($\lambda\sim$1.42--1.82\,$\mu$m), and
SK ($\lambda\sim$1.82--2.5\,$\mu$m) bands. The nominal resolutions were 500,
and 450 for the two cases, respectively.
The observations of the two targets were carried out in visitor mode on
September 13$^{th}$--14$^{th}$, 2007. Total integration times were 81\,min and
66\,min in the SH band, and 84\,min and 42\,min in the SK band on SW022550,
and SW022513, respectively.
   \begin{figure*}
   \centering
   \includegraphics[width=8cm]{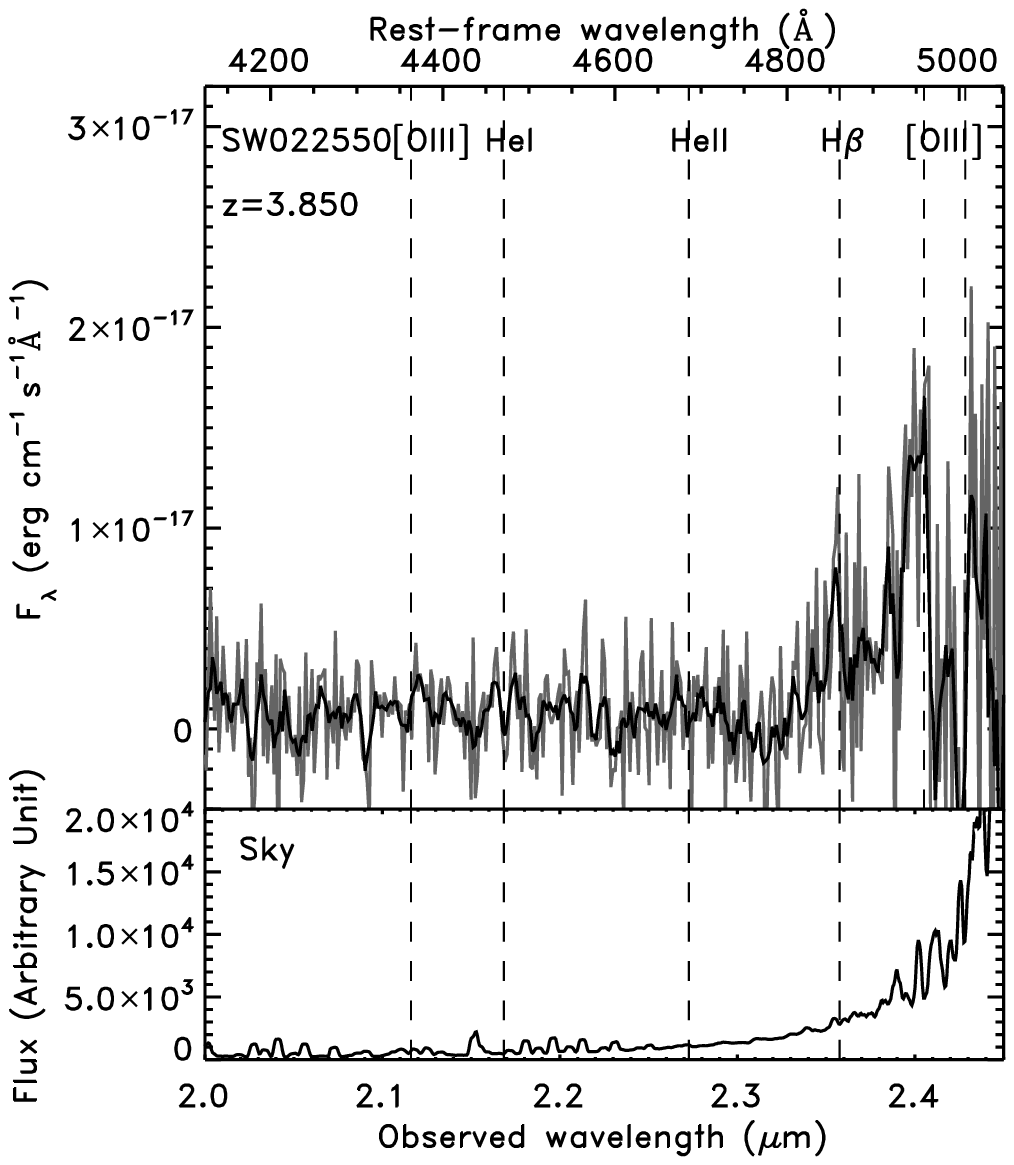}
   \includegraphics[width=8cm]{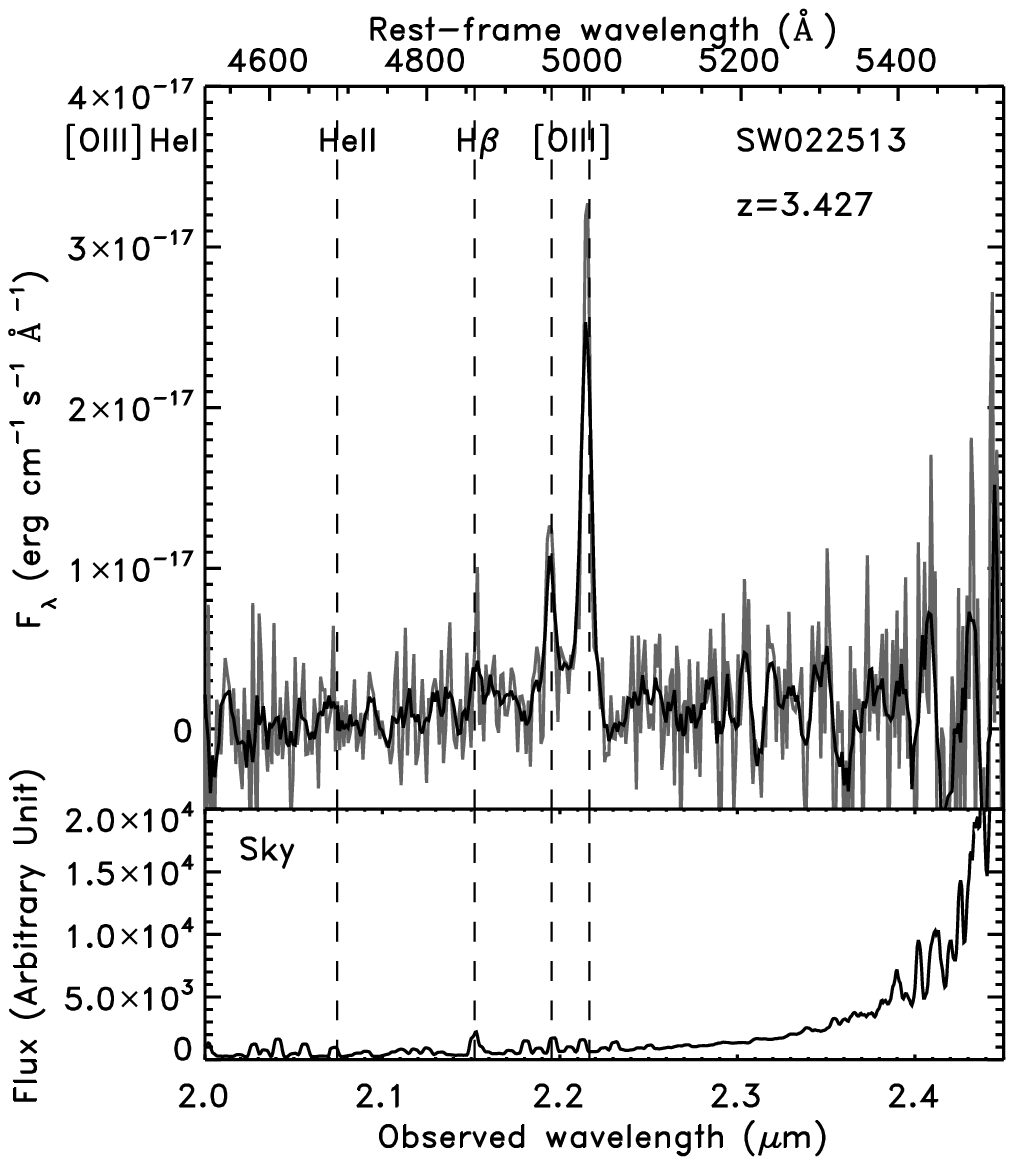}
      \caption{The top panels show the ISAAC SK spectra of SW022550 ({\it
          left panel}) and SW022513 ({\it right panel}). The grey curves
          correspond to the spectrum smoothed with a kernel of 2 pixels, and
          the black curves to the smoothed spectrum with a kernel of 10
          pixels. The bottom panels show the SK band sky spectrum. The
          location of the main expected lines are labeled.}
         \label{isaac_spe}
   \end{figure*}

Each observation was splitted in sets of 180s exposures, adopting a nodding
and jittering pattern for optimal background subtraction. The nod throw was
set to 60--80\,\arcsec, in order to avoid conflict with the reference star;
the jitter width was 10\,\arcsec.

Data reduction was carried out in the standard manner, using the IRAF
environment\footnote{The package IRAF is distributed by the National Optical
Astronomy Observatory which is operated by the Association of Universities
for Research in Astronomy, Inc., under cooperative agreement with the
National Science Foundation.}, including ABBA sky subtraction.  Wavelength
calibration was based on OH sky lines. Atmospheric extinction correction and
flux calibration were obtained with standard telluric stars, carefully selected
from the ESO database in order to have known NIR magnitudes.  For this
purpose, several B-type telluric standards were observed each night (before
and after a target was observed) at air masses within 0.1--0.2 of the air
mass of the target observations. The rebinned ISAAC SK-band spectra of the two sources are shown in
Figure~\ref{isaac_spe}, and discussed below. 

\subsubsection{ISAAC SK Spectrum of SW022550}

Neither continuum nor lines are detected in the SH-band spectrum of
SW022550. Two emission lines are detected in the SK-band spectrum at
$\lambda$23560\AA, and 24060\AA\ yielding a $z\sim$3.85 if interpreted as
\hbeta, and \oiiia, respectively. The continuum in the SK-band is not
detected and it is estimated to be $<$3.8$\times$10$^{-18}$\,\ergAcm2s\
(corresponding to 3$\sigma$). The identification of these lines is
tentative due to the high sky level at these wavelengths and the uncertain
wavelength calibration at $\lambda\geq$2.4$\mu$m. The \oiiia\ line shows a
blue tail probably due to a poor sky subtraction. At the expected location
of the \oiii\ line, the spectrum is negative, suggesting that the sky
subtraction might have been too high at these wavelengths. The
\hbeta/\lya\ flux ratio is 0.03$\pm$0.1, which is consistent with the range
of values observed in type 2 AGNs, i.e.  from 0.014 to 0.03, while type 1
AGNs are usually characterized by higher values (0.07--0.125).  Note that
because of the poor sky subtraction where the lines are observed (see bottom
panel of Figure~\ref{isaac_spe}), it is possible that a broad \hbeta\
component is also present. Assuming a FWHM of 5000\,\kms\ for a broad
\hbeta\, we estimate an upper limit to the flux
$<$16$\times$10$^{-16}$\ergcm2s. This value is significantly higher that the
predicted flux based on the measured \lya\ flux and the \hbeta/\lya\ flux
ratio observed in type 1 AGNs. Thus, we cannot rule out the presence of a
broad \hbeta\ line in the spectrum of SW022550. The \oiiia/\hbeta\ flux
ratio is 6.3$\pm$2.1. This is also closer to the ratios observed in type 2
AGNs, $\sim$2.3, than to those observed in type 1 AGNs, $\sim$0.04. The
upper limit to the \hbeta\ flux yields a \oiiia/\hbeta\ flux ratio $>$0.56,
still higher than in type 1 QSOs. In summary, in spite of its low
signal-to-noise, we conclude that the (rest-frame) optical spectrum of
SW022550 is more similar to the spectra of type 2 AGNs, than to those of
type 1 AGNs. The ISAAC spectrum shows another interesting property. The
\hbeta, and \oiiia\ lines are slightly blueshifted with respect to the
optical emission lines. Assuming that such a difference is real and that the
source is at $z$=3.867 as derived from the optical spectrum, the blueshifted
lines could be explained by outflowing gas with a velocity
$\sim$510\,\kms~\citep[see similar cases in][]{sajina08}. Such a
velocity is consistent with those assumed in shocks models~\citep{allen08}.

\subsubsection{ISAAC SK Spectrum of SW022513}

Neither continuum nor lines are detected in the SH-band spectrum of
SW022513. In the SK-band spectrum, two emission lines at
$\lambda_{obs}$=21954\,\AA, and 22160\,\AA\ are well detected, while the
continuum is not detected. We thus estimate an upper limit to the continuum
$<$2.03$\times$10$^{-18}$\,\ergAcm2s\ (corresponding to 3$\sigma$), and to
the emission lines equivalent widths. The two lines are interpreted as
\oiiipair, constraining the redshift to be $z$=3.427$\pm$0.001. A weak
emission feature is observed at the expected wavelength corresponding to the
\hbeta\ line. Since the line is affected by a bright sky line which falls
exactly at its expected location, and it is narrower than the spectral
resolution, we cannot constrain it. The \oiiia/\hbeta\ and \oiii/\hbeta\
flux ratios are, respectively, $>$0.97 and $>$2.48. These values are much
higher than those observed in type 1 AGNs (i.e. 0.04 and 0.15) and more
similar to those observed in type 2 AGNs (2.3 and 5.5) and expected from
photoionization models of the narrow line
region~\citep[NLR;][]{osterbrock89,groves04}.Thus, it is quite likely that
such a feature is the \hbeta\ line. In summary, the ISAAC spectrum of
SW022513 shows narrow emission lines with flux ratios consistent with those
of type 2 AGNs.

The lines fluxes, $W_{\lambda}$, and FWHMs of both spectra are reported in
Table~\ref{spe_tab}.The SH-band observations cover the rest-frame wavelength
range $\sim$2900--3700\,\AA\ for SW022550, and $\sim$3200--4100\,\AA\ for
SW022513. At these wavelengths, the only feature that might be detected is
the \oii\ emission line. The \oii\ line is typically associated with star
formation activity~\citep{gallagher89}, but also with AGN activity. Since
our sources are not detected in the SH-band, and an atmospheric feature is
present at the expected \oii\ location in the two sources, we cannot set any
constraints on their \oii\ emission.

From the spectra we estimate that the upper limits to the NIR continuum
in SW022513 and SW022550 are $<$33\,$\mu$Jy, and $<$61\,$\mu$Jy. These
values are about 3 and 8 times higher than the K-band measured fluxes. Thus,
the spectroscopic measurements are consistent with the broad band
photometric data.

\begin{table*}
\caption{Emission-Line Measurements}
\label{spe_tab}
\begin{center}
\begin{tabular}{l r@{$\pm$}l c r@{$\pm$}l r@{$\pm$}l r@{$\pm$}l r}
   Line   & \multicolumn{2}{c}{$\lambda_{\rm obs}$} &   $z$    & \multicolumn{2}{c}{$W_{\lambda, rest}$} & \multicolumn{2}{c}{Flux}                  & \multicolumn{2}{c}{FWHM$^a$} &    Comments \\
          & \multicolumn{2}{c}{(\AA)}               &          & \multicolumn{2}{c}{(\AA)}               & \multicolumn{2}{c}{($10^{-16}$ \ergcm2s)} & \multicolumn{2}{c}{(\kms)}   &             \\
\hline\hline
\multicolumn{11}{c}{SW022550} \\
\hline
\lyb      &  4997.4 & 0.8 &      3.876  &    41 & 3   &   1.35 & 0.11   &     935 & 180   &    Narrow  \\
\ovipair  &  5018.2 & 0.6 &      3.867  &    68 & 3   &   2.25 & 0.10   &     777 &  81   &    \nodata \\
          &  5042.9 & 0.8 &      3.858  &   152 & 4   &   5.03 & 0.14   &    1820 & 114   &    \nodata \\
\lya      &    5928 & 0.1 &      3.875  &   322 & 5   &  13.52 & 0.19   &    1041 &  13   &    Narrow  \\
\nv       &    6022 & 0.7 &      3.856  &   311 & 9   &  13.19 & 0.39   &    2531 &  99   &    Broad   \\
          &    6044 & 0.7 &      3.874  &   146 & 5   &   6.19 & 0.20   &    1354 &  74   &    Narrow  \\
\lya      &    5920 & 1.8 &      3.868  &   164 & 15  &   6.88 & 0.65   &    4060 & 157   &    Broad   \\
          &    5985 & 2.2 &    \nodata  &   496 & 23  &  \hspace*{0.5cm}20.98 & 0.96   &    8968 & 343   &    Broad   \\
\siiv     &    6778 & 4.3 &      3.852  &   833 & 37  &   7.38 & 0.33   &    4581 & 318   &    \nodata \\
          &    6820 & 2.4 &      3.861  &   281 & 22  &   2.49 & 0.20   &    2375 & 172   &    \nodata \\
\nivp     &    7242 & 2.4 &      3.873  &    41 & 4   &   1.95 & 0.18   &    1987 & 239   &    \nodata \\
\civ      &    7527 & 1.5 &      3.859  &   328 & 8   &  16.13 & 0.38   &    3690 & 101   &    Broad   \\
          &    7542 & 0.8 &      3.869  &    91 & 4   &   4.47 & 0.18   &    1235 &  88   &    Narrow  \\
\heii     &    7970 & 15  &      3.860  &    84 & 8   &   2.75 & 0.26   &    3098 & 768   &    \nodata \\
\hbeta    &  \multicolumn{2}{c}{23560}  &  3.847  &\multicolumn{2}{c}{$>$37$^b$}  &    1.4 & 0.5    & \multicolumn{2}{c}{$<$510$^c$}   &    ISAAC   \\
\oiiia    &  \multicolumn{2}{c}{24060}  &  3.852  &\multicolumn{2}{c}{$>$239$^b$} &    9.1 & 0.4    & \multicolumn{2}{c}{$<$498$^c$}   &    ISAAC   \\ 
\hline
\multicolumn{11}{c}{SW022513} \\
\hline
\hbeta    &    21532 & 5  &      3.429  & \multicolumn{2}{c}{$>$120$^b$} & \multicolumn{2}{c}{$<$2.5} & \multicolumn{2}{c}{$<$557$^c$}  &    ISAAC  \\  
\oiiia    &    21954 & 2  &      3.427  & \multicolumn{2}{c}{$>$297$^b$} &     6.1 & 0.5   &      288 & 58   &    ISAAC  \\ 
\oiii     &    22160 & 1  &      3.426  & \multicolumn{2}{c}{$>$829$^b$} &    16.9 & 0.6   &      660 & 34   &    ISAAC  \\ 
\hline\hline
\end{tabular}\\
\end{center}
{\small All measurements are based on single Gaussian fits to the emission
lines assuming a flat (in $F_{\lambda}$) continuum.}
$^a${\small Corrected for instrumental resolution.}
$^b${\small Since the continuum is not detected we estimate an upper limit
to the equivalent width, $W_{\lambda}$, assuming a 3$\sigma$ upper limit to the continuum. The
continuum is $<$3.78$\times$10$^{-18}$\,\ergAcm2s\ for SW022550, and
$<$2.03$\times$10$^{-18}$\,\ergAcm2s\ for SW022513.}
$^c${\small The FWHM is fixed to the spectral resolution because it cannot be correctly determined due to the low signal-to-noise.}
\end{table*}

\subsection{Radio data}\label{radio}

Both targets are detected in the VIMOS VLA survey at 1.4\,GHz, and at
610\,MHz~\citep{bondi03,ciliegi05,bondi07}. The measured fluxes are reported
in Table~\ref{basic_data}. In the VLA 1.4\,GHz observations, SW022513
appears extended with a major axis of 4.1\arcsec\ and a minor axis of
1.8\arcsec, corresponding to a projected linear size of
4.2\,kpc$\times$2.1\,kpc, and a position angle (PA) of
161.3\deg~\citep{bondi03}. In the other radio observations the sources are
unresolved. Assuming a power-law model $F_{\nu}\propto\nu^{\alpha_r}$ for
the radio emission, the radio spectral indeces $\alpha_{r}$ are $-$1.3 for
SW022550, and $-$0.7 for SW022513. Because of its radio spectral index,
SW022550 can be considered an ultra-steep spectrum source~\citep[USS, i.e.
$\alpha_r\leq-1$;][ and references therein]{roettgering94}. Such a steep
spectrum implies an AGN origin of the observed emission. In case of
SW022513, the spectral index is consistent with what is observed in
star forming galaxies and radio-quiet AGNs~\citep{ciliegi03}, however the
extended emission and the large radio flux strongly suggest also an AGN
origin of its radio emission. A more detailed discussion of the radio
properties of our sources is presented in \S~\ref{radio_disc}.

\subsection{X ray data}\label{xray}

The two targets are also covered by 20\,ks observations with \xmm\ from the
\xmm\ Medium Deep Survey~\citep[XMDS;][]{chiappetti05,tajer07}. Only
EPIC/MOS data are available as both sources fall in a bad column of the
EPIC/pn detector. SW022513 is at 4.5\arcmin\ from the aimpoint and
marginally detected with 29$\pm$12 total counts (12$\pm$8 at 0.5--2\,keV,
and 17$\pm$9 at 2--10\,keV). The hardness ratio is HR=0.18$^{+0.7}_{-0.3}$,
which corresponds to an effective gas column density
\nh$\simeq$(1$^{+4}_{-0.5})\times$10$^{24}$\,\cm2\ assuming a power-law
model with photon index $\Gamma$=1.7, and Galactic and intrinsic
photo-electric absorption with \nh$^{Gal}$=
2.61$\times$10$^{20}$\,\cm2~\citep{dickey90}. More details on the X ray data
and derived quantities can be found in~\citet{chiappetti05},
and~\citet{tajer07}.  SW022550 is at $>$11\arcmin\ from the aim-point and is
not detected, thus we can only give an upper limit to its X-ray flux.  The
X-ray fluxes of both sources are reported in Table~\ref{basic_data}.

Based on these X ray measurements, the estimated broad and hard X ray
luminosities of SW022513 are $\sim$6$\times$10$^{44}$\,\ergs, and
5$\times$10$^{44}$\,\ergs, respectively. The absorption-corrected broad and
hard X ray luminosities are 11$\times$10$^{44}$\,\ergs, and
6$\times$10$^{44}$\,\ergs. For SW022550, we estimate an upper limit to the
absorbed luminosities in the broad and hard bands of
2.0$\times$10$^{45}$\,\ergs, and 1.8$\times$10$^{45}$\,\ergs, respectively.
Based on the estimated X ray luminosity of SW0221513 and the relationship
between the \oiii\ and the hard X ray luminosity in type 2
AGNs~\citep{mulchaey94,netzer06}, we expect an \oiii\ luminosity of
3--9$\times$10$^{42}$\,\ergs, instead of 2$\times$10$^{44}$\,\ergs\ as
observed (see Table~\ref{tab_lum}). Such a discrepancy might be due to the
non validity of the X ray/\oiii\ relationship for our sources, to an
underestimation by a factor of 100 of the X ray luminosity, or to the
contamination from star formation to the \oiii\ flux. Probably, all these
factors play a role in explaining this discrepancy, but the only one that
could explain such a large difference is the underestimation of the
intrinsic X ray luminosity. A higher X ray luminosity would also be required
to obtain a MIR/X-ray luminosity ratio consistent with those observed in
other AGNs (see \S~\ref{comparison}). Based on these considerations, it is
quite probable that SW022513 is a Compton-thick QSO. Since SW022550 is even
fainter in the X-rays, and more luminous in \oiii\ emission and in MIR
luminosity, it is also quite plausible that SW022550 is a Compton-thick QSO.

\begin{figure*}[ht!]
  \begin{center} 
 \includegraphics[keepaspectratio='true',scale=0.7]{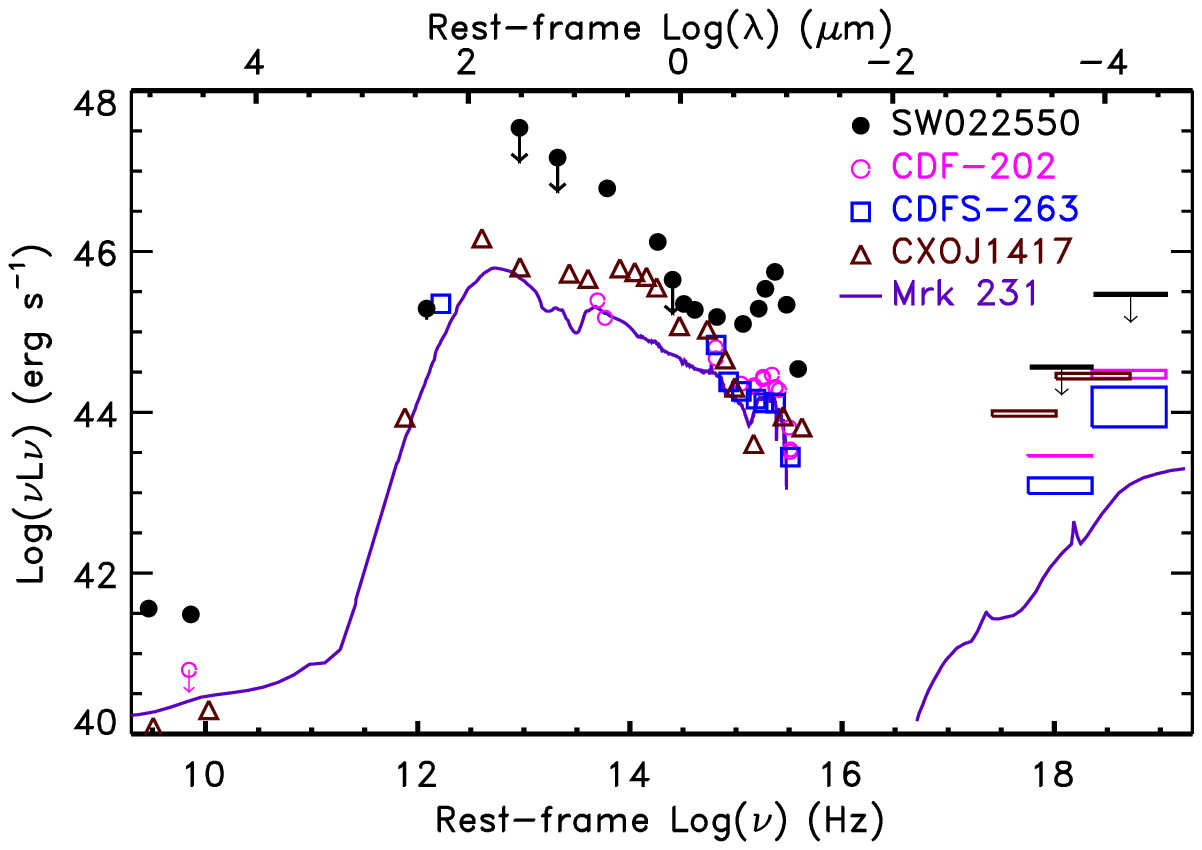}
 \includegraphics[keepaspectratio='true',scale=0.7]{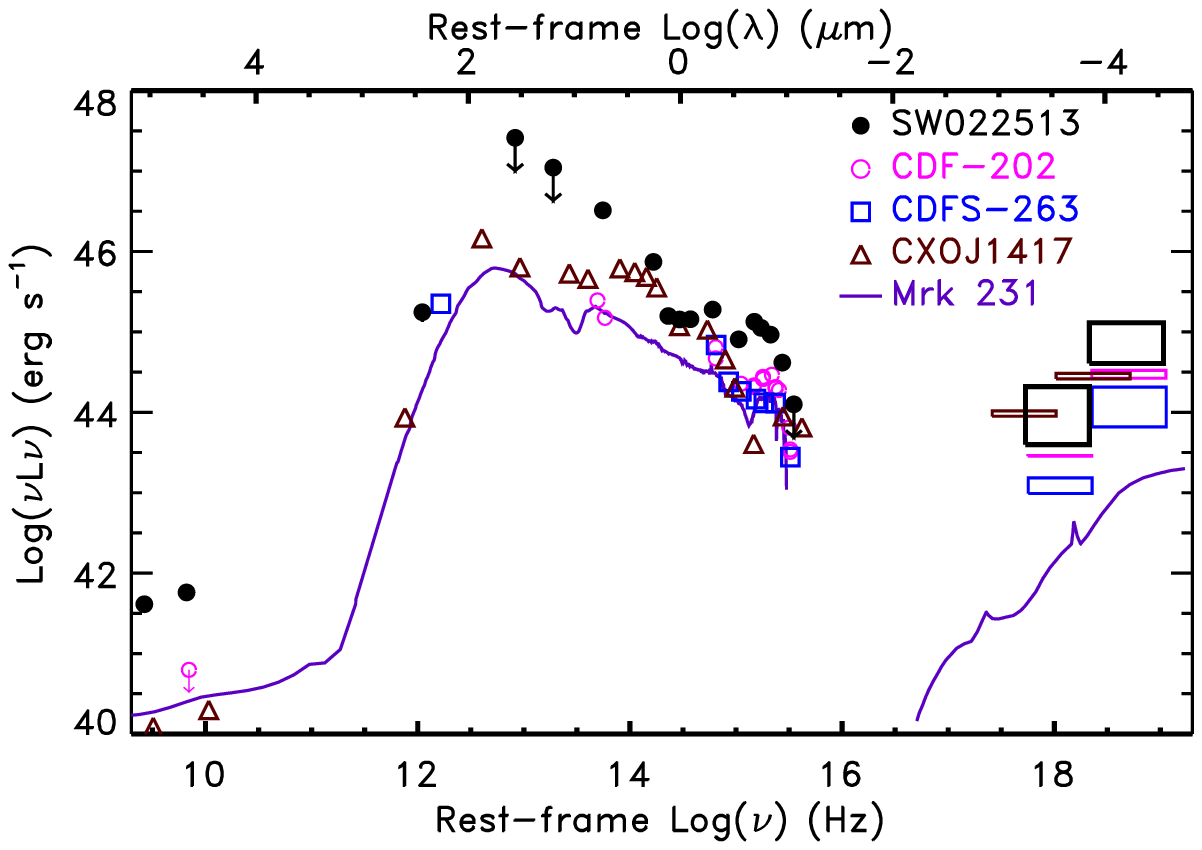}
  \end{center}
\vspace*{-0.5cm}
      \caption{{\small Rest-frame radio-X ray spectral energy distributions
of the selected targets (black full circles), SW022550 ({\it left panel}),
and SW022513 ({\it right panel}) compared with the SEDs of X ray detected
heavily absorbed AGNs from the literature, CDF-202 at $z$=3.7~\citep[magenta
circles;][]{norman02,miller08}, CDFS-263 at $z$=3.66~\citep[blue
squares;][]{mainieri05}, CXOJ1417 at $z$=1.15~\citep[brown
triangles;][]{lefloch07}, and Mrk\,231 at $z$=0.042 (purple solid line). The
rectangles represent the soft and hard X ray fluxes and associated
uncertainties. Downward arrows represent 5$\sigma$ upper limits. }}
         \label{seds}
   \end{figure*}

The full radio-Xray SEDs of the two targets are shown in Figure~\ref{seds},
and all the available fluxes are listed in Table~\ref{basic_data}.

   \begin{figure*}
   \centering
   \includegraphics[width=18cm]{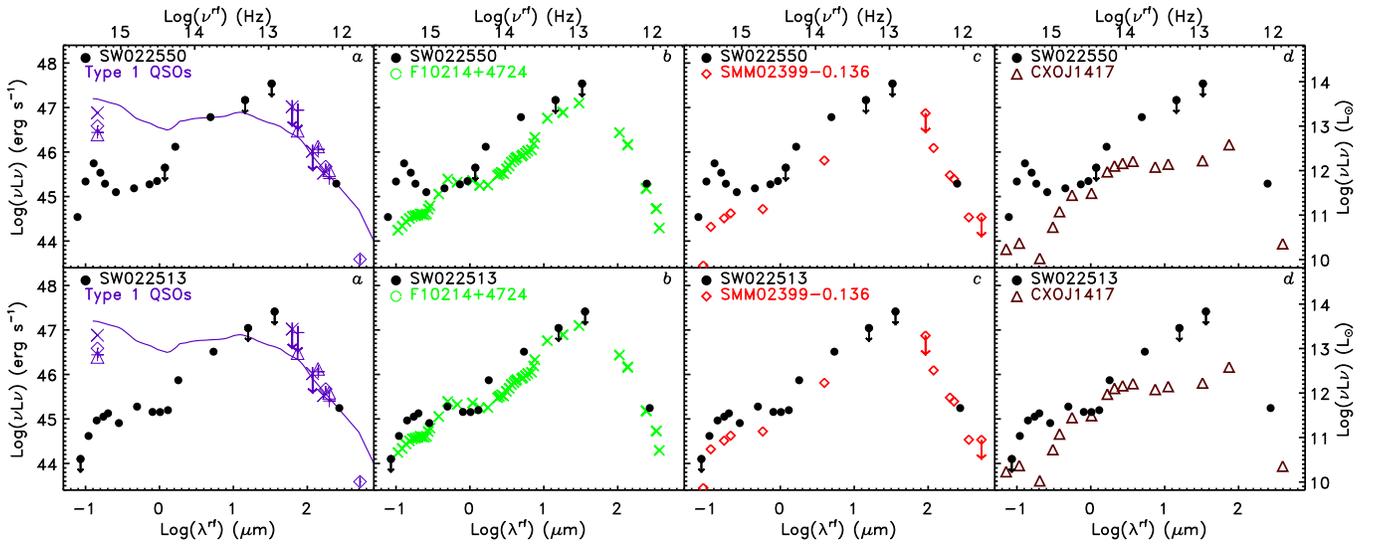}
      \caption{Rest-frame UV-mm SEDs of the selected targets (black full
circles), SW022550 ({\it top panels}), and SW022513 ({\it bottom panels}),
compared with the SEDs of other mm-detected AGNs, ($a$) a median type 1 QSO
template normalized at the mm luminosity~\citep{elvis94a}, and 4 mm-detected
type 1 QSOs with FIR excess~\citep[purple symbols]{wang07z6}, J033829.31+002156.3 (purple plus
signs), J075618.14+410408.6 (purple triangles), J092721.82+200123.7 (purple
diamonds), and J104845.05+463718.3 (purple crosses), ($b$)
F\,10214+4724~\citep[green crosses;][]{rowan-robinson93,teplitz06},
($c$) SMM\,02399$-$0136~\citep[red diamonds;][]{ivison98}, 
and ($d$) CXOJ1417~\citep[brown triangles;][]{lefloch07}.}
         \label{multi_seds}
   \end{figure*}

   \begin{figure*}[ht!]
   \centering
   \includegraphics[width=8cm]{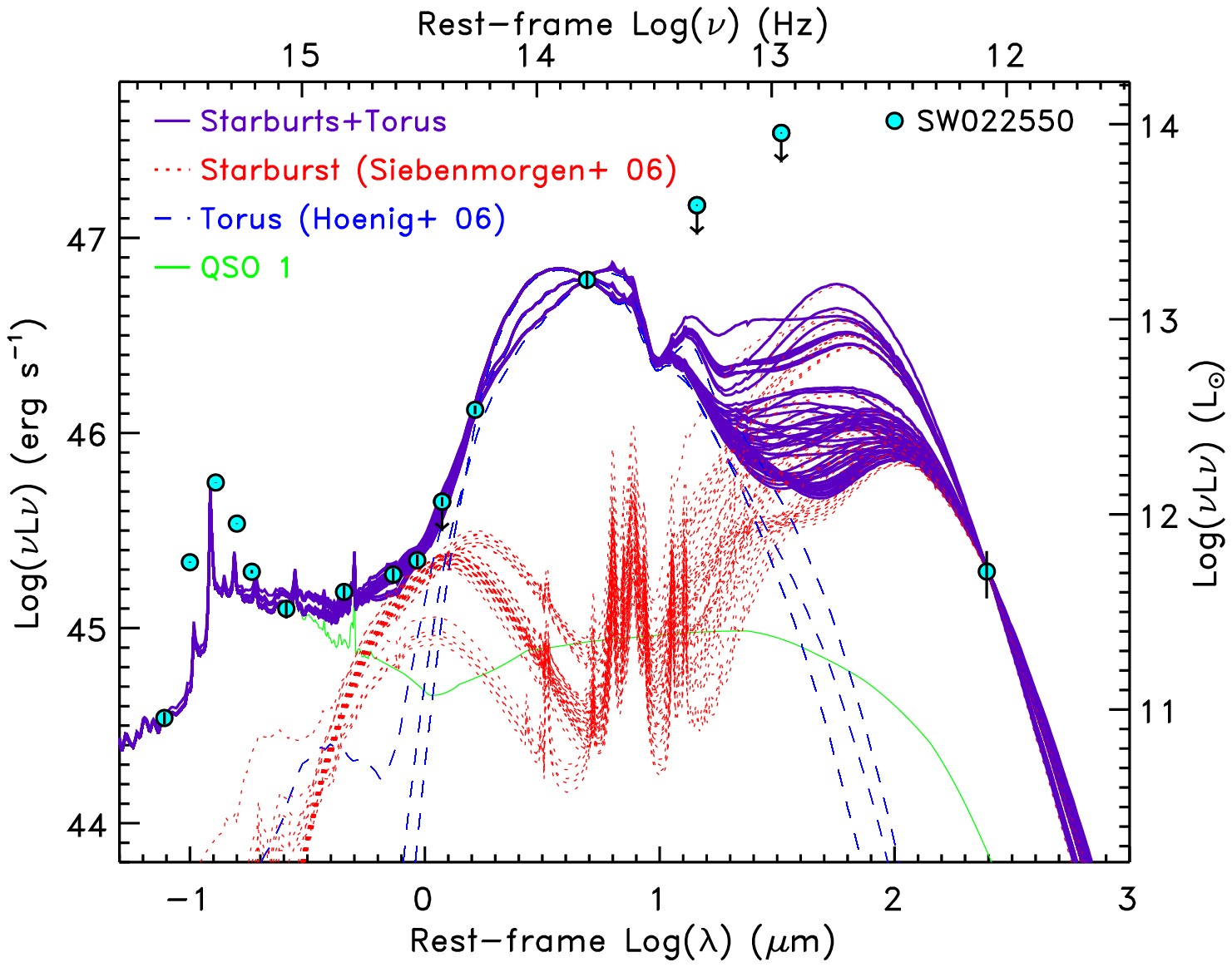}
   \includegraphics[width=8cm]{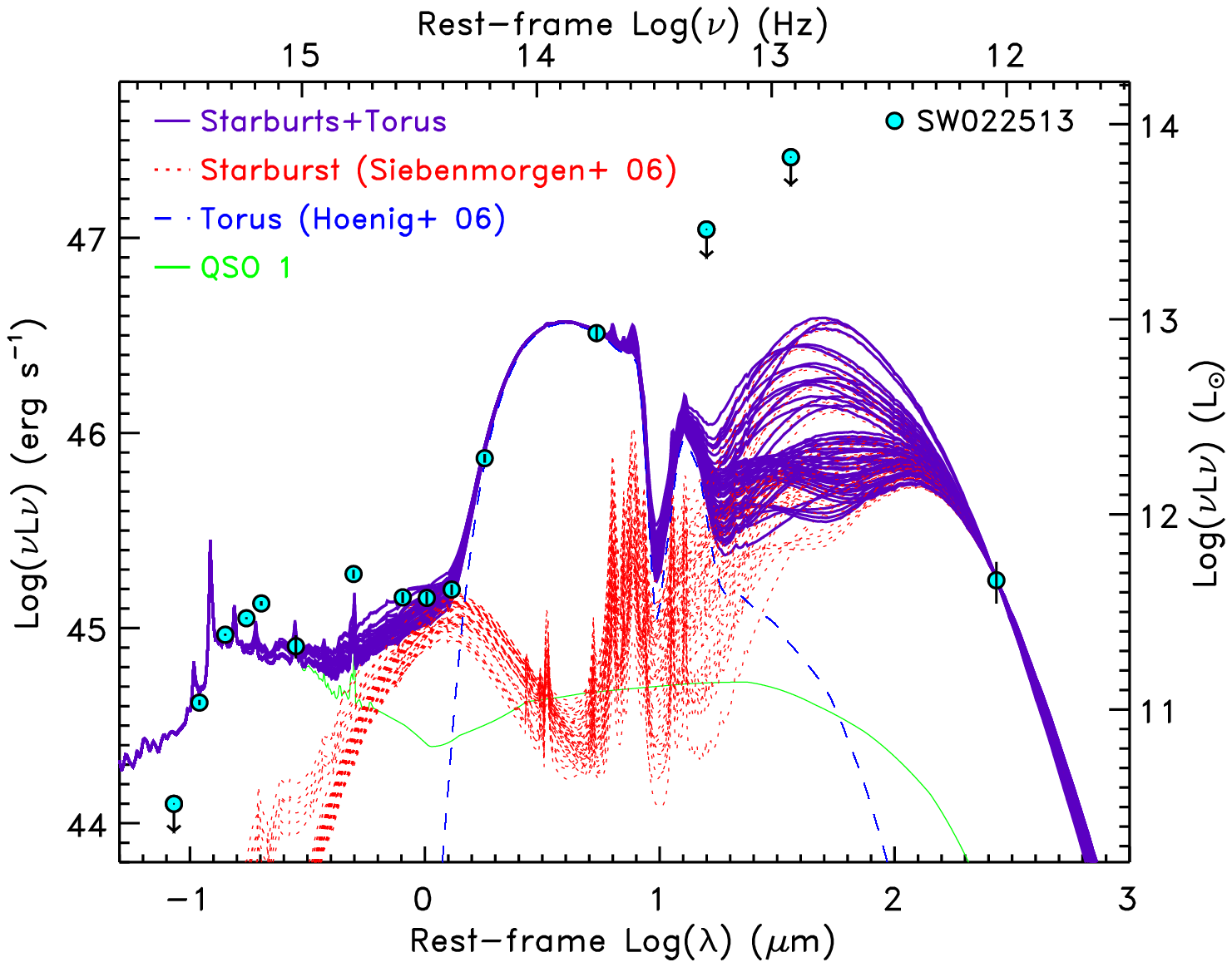}
      \caption{Rest-frame UV-mm spectral energy distributions of the
selected targets (cyan full circles), SW022550 ({\it left panel}), and
SW022513 ({\it right panel}). The red dotted curves represent the starburst
models normalized at the observed 1.2\,mm flux from the library
in~\citet{siebenmorgen06}. The dashed blue curves represent torus models
from~\citet{hoenig06} normalized at 24\,$\mu$m. The green solid line is a
type 1 AGN template normalized at the observed J-band (1.25\,$\mu$m) flux.
The solid purple curves represent the total of the starburst, torus, and type
1 AGN templates which yield the 50 best $\chi^2$, where the $\chi^2$ is
computed using 7 data points from the J-band to the 24\,$\mu$m,
and do not overpredict the observed fluxes and upper limits at
FIR wavelengths (see text for more details). The models are used
to estimate the luminosities reported in Table~\ref{tab_lum}.}
         \label{fir_seds}
   \end{figure*}

\section{Multi-wavelength spectral energy distributions}

The SEDs of SW022550, and SW022513 are shown in Figures~\ref{multi_seds}
and~\ref{seds}. Both sources are characterized by large MIR fluxes
(F$_{24\mu m}$$\simeq$2--3\,mJy), show red power-laws in the NIR
(1--5$\mu$m in the rest-frame; $\alpha_{IR}$$<$$-$2.5, where F$_{\nu}$$\propto$
$\nu^{\alpha_{IR}}$), and large IR/optical flux ratios ($Log(\nu
F_{24\,\mu m}/\nu F_{z})$=1.5). All these properties are signatures typical
of obscured QSOs~\citep{weedman06a,polletta08a}. 

\subsection{Comparison with other high-$z$ QSOs}\label{comparison}

We compare the UV-mm SEDs of SW022550, and SW022513 with those of well
studied mm-detected non radio-loud AGNs at high-$z$ from the literature in
Figure~\ref{multi_seds}. The vast majority of mm-detected QSO at high-$z$
are unobscured or type 1 QSOs with MIR SEDs consistent with the median
template of unobscured QSO~\citep[e.g.][]{hines06}. We represent their SEDs
in panel $a$ of Figure~\ref{multi_seds} with the median QSO template
by~\citet{elvis94a}. The majority of these type 1 QSOs with strong mm
fluxes, i.e. $\gtrsim$3\,mJy, as our sources, show a significant excess in
the far-IR (FIR) with respect to the optical emission predicted by the
median QSO template~\citep{wang07z6,hao08}. The SEDs of 4 QSOs with a FIR
excess are also shown in panel $a$ of Figure~\ref{multi_seds}. Our sources
show similar mm luminosities, but 10--100 times lower optical fluxes,
consistent with being obscured. In Figure~\ref{multi_seds}, we also compare
our sources SEDs with those of 3 mm-detected obscured QSOs with available
MIR data, (1) the lensed system \object{F\,10214+4724}, a Seyfert 1 galaxy
at $z$=2.86, but that is also heavily absorbed in the X rays~\citep[panel
$b$;][]{ivison98,alexander05c}; (2) \object{SMM\,02399$-$0136}, a type 2 QSO
at $z$=2.803~\citep[panel $c$;][]{ivison98}; and (3) \object{CXO GWS
J141741.9+522823} (CXOJ1417 hereinafter) at $z$=1.15~\citep[panel
$e$;][]{lefloch07}. With the only exception of CXOJ1417 that is the least
luminous sources of the sample, all the remaining cases are characterized by
mm luminosities consistent with those of our sources. All the obscured AGNs
shown in panels $b$--$d$ in Figure~\ref{multi_seds} show flat NIR SEDs with
similar luminosities. Since the maximum of stellar light in $\nu F_{\nu}$
and the minimum of AGN light are at NIR wavelengths~\citep[1--1.6$\mu$m in
the rest-frame;][]{sawicki02,sanders89}, it is at these wavelengths that we
can expect the maximum contribution to the optical-IR SED from the host
galaxy. The similarity of the observed NIR SEDs might thus be explained by a
significant contribution from stellar emission. All SEDs are characterized
by red MIR colors, consistent with hot dust thermal emission, but such a
component is more luminous and redder in our sources than in the sources
from the literature. Our sources are also systematically more luminous at
optical wavelengths, and this is mainly due to the strong emission lines. In
summary, our targets have SEDs that are intermediate between those of type 1
QSOs ({\it panel a}), and those of obscured QSOs ({\it panels b-d}). In
absence of high spatial resolution data and MIR spectroscopy, we cannot
determine the origin of the observed differences. Assuming that the
intrinsic AGN SED is the same in all sources, the redder MIR SED and the
higher optical flux imply a higher AGN luminosity and more extinction in our
sources than in the mm-detected obscured QSOs from the literature.

In Figure~\ref{seds}, we compare our sources radio-X ray SEDs with those of
high-$z$ AGNs with similar multi-wavelength coverage. We include the
mm-detected sources CDFS-263, and CXOJ1417 because they are also detected at
X ray wavelengths. CXOJ1417 is also detected at radio wavelengths. We also
include CDF-202, a X ray detected heavily absorbed type 2 QSO at
$z$=3.7~\citep{norman02}, and \object{Mrk\,231}, a BAL QSO which is heavily absorbed
in the X rays and is hosted by a powerful starburst galaxy. We do not
consider F\,10214+4724 because its X ray spectrum is likely dominated by
starburst emission~\citep{alexander05c}. The absorption-corrected hard X ray
luminosities of CDFS-263, CXOJ1417, and CDF-202 are 7.6$\times$10$^{44}$,
2.4$\times$10$^{44}$, and 3.3$\times$10$^{44}$\,\ergs, respectively. The X
ray luminosity of SW022513 is $\sim$6$\times$10$^{44}$\,\ergs. Unless, the X
ray luminosity of SW022513 is largely underestimated (see \S~\ref{xray}),
the current estimate is consistent with those measured in the literature
sources. In addition to being more luminous at all wavelengths, our
sources show a redder MIR emission than the CDFS sources.  Interestingly,
such a red MIR emission is also observed in other heavily obscured QSOs at
high-$z$ which are also characterized by higher MIR/X ray luminosity ratios
($\sim$13--40) than classical AGNs~\citep{polletta08a}. For classical AGNs
the MIR/X ray luminosity ratios ($\nu L_{6 \mu m}/L_{2-10 keV}$) range from
about 0.3 to 13~\citep{polletta07,polletta08a}, but for SW022513 and
CXOJ1417 this ratio is about 60. Such a prominent MIR component might imply
either that dust heating is more efficient, that the dust covering factor is
particularly high, or that the intrinsic X-ray luminosity is underestimated.

\subsection{SED modeling and star formation rate}\label{model}

Our targets have optical and MIR properties typical of obscured QSOs, but
their large mm fluxes suggest that they are also experiencing intense star
forming activity. Indeed, emission from AGN-heated circumnuclear dust is
expected to peak around 10--30\,$\mu$m and not to contribute significantly
in the FIR~\citep[see e.g.][]{granato04,hoenig06,fritz06}. A powerful
starburst provides a more likely explanation, but such a hypothesis cannot
be confirmed because of the lack of detections at rest-frame wavelengths
$\lambda$=6--250\,$\mu$m, contrarily to the cases for F\,10214+4724 and
CXOJ1417. Although current data do not constrain the origin of the mm
emission, AGN model predictions and the similarity with F\,10214+4724 and
CXOJ1417 favor a starburst origin.

Assuming that the mm flux is due to a starburst component and that the MIR
flux is mainly due to the AGN, we model the IR (2--1200\,$\mu$m) observed
SED of our targets by combining a starburst and a torus model. To model the
starburst component, we use a library of 7208 starburst
models~\citep{siebenmorgen06}. To model the AGN component, we use a set of
44 clumpy torus models with various inclinations from~\citet{hoenig06}.
These models were used to fit the MIR SEDs and spectra of a sample of
MIR-luminous obscured QSOs with available MIR spectra~\citep{polletta08a}.
The starburst component is normalized at the observed 1.2\,mm flux, and the
torus models at the observed 24\,$\mu$m flux. The sum of the starburst and
torus models is then compared to the SED of our targets. Since none of the
models is able to reproduce the optical data, and since the optical data
indicate that there is a contribution from strong emission lines, typical of
AGNs, an additional component is added to the starburst+torus models
represented by an AGN template. For this component we use a type 1 QSO
template~\citep{hatziminaoglou05a,polletta07} because it reproduces well the
observed optical broad-band data. The type 1 QSO template is normalized at
the J-band (1.25\,$\mu$m) observed flux. Note that such a normalization
corresponds to $\sim$1.4\% of what would be obtained if we normalized the
type 1 QSO template at the observed 24\,$\mu$m flux. Such a small fraction
implies that the AGN optical light might be suppressed by more than
70\%, which corresponds to a foreground extinction of \av=4.6. Since this
component is not reddened, its origin is more likely due to a small (1.4\%)
fraction of nuclear light that escapes without suffering obscuration, or to
nuclear scattered light~\citep[see e.g.][]{zakamska06}.

After combining all the starbursts models, with all the torus models and
with the type 1 QSO template, we then reject all combinations that
overpredict the fluxes at 70, and 160$\mu$m. This criterion rejects about
40\% of all models. We then select 50 models with the lowest $\chi^2$,
computed by comparing the predicted and observed fluxes in 7 bands, from
1.25\,$\mu$m to 24\,$\mu$m. The best 50 models are shown in
Fig.~\ref{fir_seds}. The choice of 50 models yields a wide enough range of
solutions to characterize all acceptable models and the luminosities
produced by each component. A higher number of models yields either poor
fits or does not increase significantly the range of SEDs already
represented by the 50 best models.

The preferred starburst models contain an old stellar population component,
correspond to starbursts of intermediate sizes, i.e. 3\,kpc, and show a wide
range of ratios, $\sim$40--90\%, of the OB stars luminosity with respect to
the total luminosity. The preferred torus model for both sources,
defined as the most often chosen among the best 50 models, corresponds to a
torus with axis at only 15\deg\ from the line of sight with an additional
cold dusty absorber along the line of sight. Such a model explains the
prominent MIR emission as radiation from hot dust in the torus inner walls,
and the red MIR colors as extinguished torus radiation by a cold
absorber~\citep[for details on the AGN models see][]{polletta08a}.

From the set of acceptable models we derive the AGN and the starburst
bolometric (0.1--1000\,$\mu$m) luminosities, and the IR (8--1000\,$\mu$m)
starburst luminosity. The derived luminosities for all acceptable models are
reported in Table~\ref{tab_lum}.

Assuming that the estimated IR luminosity is produced by a starburst, the
derived SFRs, given by SFR(\msun/yr) =
L(IR)/(5.8$\times$10$^9$\lsun)~\citep{kennicutt98}, are
$\sim$500--3000\,\msun/yr for both sources. The estimated SFRs are
consistent with those estimated in
other radio-quiet obscured QSOs at
high-$z$~\citep[e.g][]{ivison98,efstathiou06,stevens05,polletta08a,sajina07b}.

\begin{table}
\begin{minipage}[t]{\columnwidth}
\caption{SW022550 and SW022513 luminosities}
\label{tab_lum}
\centering       
\renewcommand{\footnoterule}{}
\begin{tabular}{llcc}
Parameter\footnote{\small L(IR) is the 8--1000\,$\mu$m luminosity in \lsun\
derived by integrating the model under Method. L$_{1.4\,GHz}$ in the
monochromatic radio luminosity at 1.4\,GHz in the rest-frame in W
Hz$^{-1}$. L(H) is the monochromatic luminosity in \lsun\ at 1.6\,$\mu$m
in the rest-frame derived from the starburst models. M$_{*}$ is
the stellar mass derived from L(H) assuming the average L(H)/M$_*$ ratio
derived from the high-$z$ radio galaxy sample in~\citet{seymour07}.
L(6$\mu$m) is the rest-frame luminosity at 6$\mu$m in \lsun\ derived from
the torus models. L$^{SB}_{bol}$ is the 0.1--1000\,$\mu$m luminosity derived
from the starburst models in \lsun. L$^{AGN}_{bol}$ is the 0.1--1000\,$\mu$m
luminosity derived from the torus models in \lsun.}
        &   Method\footnote{\small The SK07 starburst models are from the library
in~\citet{siebenmorgen06}. The H06 torus models are from~\citet{hoenig06}. 
PL stands for power-law model with spectral index
$\alpha_r$, $F_{\nu}\propto \nu^{\alpha_r}$, where $\alpha_r$=$-$1.3 for
SW022550, and $-$0.7 for SW022513. }
            & SW022550   &  SW022513  \\
\hline\hline     
Log(L(IR))& SK07 Starburst           & 12.5--13.3 & 12.5--13.2  \\
Log(L$_{1.4\,GHz}$) & PL             & 25.50      &   25.38     \\
Log(L(H))  & SK07 Starburst          & 11.2--11.9 &  11.3--11.6 \\
Log(M$_{*}$) & L(H) (S07)            & 11.2--11.9 &  11.3--11.6 \\
Log(L(6$\mu$m)) & H06 Torus          & 13.1--13.2 &  12.9--12.9 \\
Log(L$^{SB}_{bol}$) & SK06 Starburst & 12.6--13.4 &  12.4--13.2 \\
Log(L$^{AGN}_{bol}$) & H06 Torus     & 13.4 &  13.1 \\
\hline
\end{tabular}
\end{minipage}
\end{table}

\subsection{The origin of the radio luminosity}\label{radio_disc}

Among the various feedback mechanisms that have been suggested as star
formation regulators in high-$z$ galaxies, the radio feedback mode is
probably the most popular and successful in reproducing numerous observables
through simulations~\citep{silk98,croton06b,cattaneo07}. Because of
the predicted short timescale associated with the radio activity and of the
difficulty of identifying AGN-driven radio activity in non radio-loud
objects, there is only limited evidence of radio feedback at play on
star-formation in AGNs, especially at high-$z$, and at high luminosities.
The best supporting observations are the detection of AGN-driven outflows in
high-$z$ radio-galaxies~\citep{nesvadba06,nesvadba07a,nesvadba07b}, and of
moderate AGN-driven radio activity in high-$z$ \spitzer-selected starburst
and AGNs~\citep{sajina07a,polletta08b}. Finding further evidence of
AGN-driven radio activity, especially in massive high-$z$ star forming
galaxies, would provide additional support for the radio feedback as
quenching mechanism. Here, we analyze the origin of the radio emission in
our two sources.

The measured radio fluxes of SW022550, and SW022513 correspond to 1.4\,GHz
rest-frame luminosities of 10$^{25.50}$\,\WHz and 10$^{25.38}$\,\WHz,
respectively, or SFRs$\sim$6000\,\msun/yr, if powered by star formation.
These high luminosities would imply the presence of AGN-driven radio
emission in the local universe~\citep[an AGN is classified radio-loud if
$P_{1.4 GHz}>10^{23.5}$\,\WHz, corresponding to a
SFR~$\simeq$100\,\msun/yr;][]{condon92}. However, at high-$z$, intense
starburst episodes with SFRs of thousands of \msun/yr can be expected. It is
thus not straightforward to infer an AGN origin of the radio emission in
sources at high-$z$ based on the estimated radio luminosities.
Although a broad multi-wavelength coverage is available for both sources,
some traditional diagnostic methods would not be reliable in probing an
AGN origin of the radio emission. The commonly diagnostic based on 
large radio/optical flux ratios cannot be applied because our sources suffer from
heavy optical obscuration. Another diagnostic is provided by the
radio/infrared flux ratio, which is based on the well established
far-IR/radio correlation~\citep{condon92}. 
The average FIR-to-radio emission ratio $q$ =
log(L(FIR)/3.75$\times$10$^{12}$\,W)$-$log(L$_{\nu,1.4GHz}$/W\,Hz$^{-1}$) is
equal to 2.34 in local star forming galaxies~\citep{yun01}, where
L(FIR)=L$_{40-120\mu m}$, and L(FIR)/L(IR)=0.56$\pm$0.1 for the starburst
models used in this work. Slightly lower values have been found in
sources at higher $z$, $q$=2.07 in $z\sim$2 SMGs~\citep{kovacs06}, and
$q$=2.21$\pm$0.02 in radio-quiet MIR selected $z\sim$1--2.5
AGNs~\citep{sajina08}. As the AGN contribution to the radio emission
increases, $q$ decreases. For example, in radio loud AGNs $q$ is typically
close to zero or negative. The $q$ values derived for our sources 
using the FIR luminosities obtained from the starburst models (see
Table~\ref{tab_lum}) are $\sim$0.8--1.6 for both sources. These low values
favor an AGN origin for the observed radio emission. This is also
supported by the steep radio spectral index in SW022550 and the extended
radio size of SW022513.

In summary, we find some evidence of an AGN origin for the radio emission of
our two sources. The associated radio activity is moderate, as it is
only slightly in excess of what is typically observed in radio quiet quasars
and much less than in radio loud quasars.

\subsection{AGN and starburst contribution to the bolometric luminosity}

The SEDs of our sources indicate that their MIR emission is dominated by
AGN-heated dust. Using the best torus models described in \S~\ref{model}
and shown in Fig.~\ref{fir_seds}, we estimate a lower limit to the AGN
bolometric luminosity by integrating the torus model in the
0.1--1000\,$\mu$m wavelength range. The derived AGN bolometric luminosities,
also reported in Table~\ref{tab_lum}, are 10$^{13.4}$\,\lsun, and
10$^{13.1}$\,\lsun\ for SW022550 and SW022513, respectively. To derive the
starburst bolometric luminosity, we do the same integration but using only
the starburst models. Based on these approximations, the AGN contribution to
the bolometric luminosities is $\sim$50--87\% in SW022550, and 43--83\% in
SW022513. The broad range of values is due to the uncertainty on the
starburst IR luminosity. The AGN contribution to the system bolometric
luminosity is thus at least 40\%, and it can be more than 80\%.

\subsection{Host stellar luminosity}\label{masses}

The stellar emission, in $F_{\nu}$, from a host galaxy peaks in the NIR,
typically at 1.6\,\micron\ in the rest-frame (or H-band)~\citep{sawicki02}.
The ratio between NIR luminosity and stellar mass is characterized by little
dispersion, especially when similar star formation histories are assumed.
The estimated stellar masses can be 30\% lower in case of younger stellar
populations. Thus, we can assume that the NIR luminosity is a proxy of the
stellar mass or luminosity of the host galaxy.  Since in our objects the AGN
optical and NIR emissions are obscured, we can directly estimate the host
NIR luminosity from the observed SED. Indeed, the best-fit SED models show
that the NIR emission is dominated by stellar light.  We estimate the host,
and thus stellar, H-band host luminosities using the acceptable starburst
templates described above. From the H-band luminosity, we estimate the
stellar mass using the average L(H)/M$_*$ ratio derived for a sample of
high-$z$ radio galaxies by~\citet{seymour07}, Log(M$_*$) =
Log(L(H))$-$0.1$\pm$0.1, where M$_*$ is the stellar mass in \msun\ and L(H)
is the monochromatic H-band luminosity in \lsun. Note that this relationship
assumes a~\citet{kroupa01} initial mass function (IMF). The estimated
stellar masses are 1.3--6.5$\times$10$^{11}$\,\msun\ for SW022550, and
1.5--2.9$\times$10$^{11}$\,\msun\ for SW022513. The estimated H-band
luminosities and stellar masses are reported in Table~\ref{tab_lum}. Because
of the uncertainty in separating the host and the AGN contributions to the
H-band luminosity, the stellar masses should be considered as upper limits
to the true stellar masses. The estimated masses are of the same order as
those of the most massive systems at high-$z$~\citep[see
e.g.][]{seymour07,berta07a}, and are thus among the most massive objects at
$z$$\sim$3--4.

\section{Discussion}\label{discussion}

\subsection{Dust obscuration: geometry and distribution}

The two sources analyzed in this work exhibit properties usually observed in
type 2 AGNs. Their narrow line dominated spectra, low optical/IR flux
ratios and faint X ray fluxes imply that they are obscured, but both sources
show a blue optical continuum which is unexpected by the standard AGN
unification model~\citep{antonucci93}, but not unusual for obscured
AGNs~\citep[see e.g. SW104409;][]{polletta06}. Moreover, SW022550 shows some
broad components in its optical spectrum. In addition, both sources also
exhibit a prominent MIR emission suggesting that the hottest dust component,
likely at the dust sublimation radius, is in part visible (see preferred
torus models in \S~\ref{model}). These hybrid properties suggest that a
fraction of the nuclear light is visible. Since the SEDs of our sources
cannot be simply reproduced by extinguishing the emission of a type 1 QSO, a
more complex model is required to explain their properties. A proper
modeling, that takes into account various geometries, dust properties, the
intensity and the spectrum and luminosity of the heating source, and a
radiative transfer treatment would be necessary to fully explain their
hybrid properties. However, the uncertainty on determining the spectrum of
the AGN component due mainly to the contribution from the host galaxy, and
the lack of a MIR spectrum, would yield degeneracies in the models. 
Thus, we limit our discussion to a qualitative comparison with some of the
most popular models of obscuring dust in AGNs from the literature.  A
possible scenario is a clumpy dust distribution which produces a partial
covering of the nucleus~\citep{nenkova02,hoenig06}. This scenario is quite
unlikely if the optical radiation comes from the nuclear region, and it is
thus much smaller than the dust sublimation radius. An alternative to an
unobstructed line of sight to the nucleus is having a scattering medium,
perhaps associated with the NLR. The scattering medium can produce the blue
optical light, and broad emission lines. If dust is mixed with the
scattering medium, this might be responsible for the large MIR excess, while
the nuclear light is obscured by the dusty torus~\citep{schweitzer08}. 
Spectropolarimetric observations to measure the scattered component are
planned for SW022550 to investigate the latter scenario.

\subsection{SMG--QSO evolutionary link}

Our targets show simultaneous powerful starburst and AGN activity. They are
thus good candidates for transitioning objects between the starburst and the
QSO stage predicted by the most popular evolutionary
models~\citep{sanders88,alexander05a}. Here, we consider their properties to
make predictions on their evolution.

From the AGN bolometric luminosity (see Table~\ref{tab_lum}), and assuming
that the AGN emission is Eddington limited ($L_{bol}=L_{Edd}$), we derive a
lower limit to the BH masses of SW022550 and SW022513 of $\simeq 7.4\times
10^8$ \Msun\ and 3.7$\times 10^8$ \Msun, respectively. Since most of the
quasars accrete below their Eddington limit~\citep{mclure04}, it would be
more realistic to assume a lower Eddington ratio, but this would imply even
higher BH masses. Note that these estimates are not quite accurate since we
are estimating the AGN bolometric luminosity from the reprocessed thermal
emission which is highly dependent on the properties of the reprocessing
dust~\citep[see e.g.][]{marconi04}.  Assuming an accretion efficiency of
10\%, the derived accretion rates are $\sim$17~\Msun\ yr$^{-1}$ for SW022550
and 8.5~\Msun\ yr$^{-1}$ for SW022513.

Assuming the estimated BH masses and the stellar masses derived in
\S~\ref{masses}, we find that these sources lie on the local
M$_{BH}$-M$_{bulge}$ relationship~\citep[e.g.][]{marconi03}. Thus they
are more similar to other obscured QSOs at high-$z$, than to SMGs with AGN
activity or type 1 AGNs which are both offset from the local M$_{BH}$-M$_{bulge}$
relationship~\citep{alexander08,polletta08b,coppin08b}. In order to keep these objects
on the local M$_{BH}$-M$_{bulge}$ relation as they evolve, the on-going star
formation process and BH growth will have to stop, or continue with
analogous rates (SFR$\equiv$\.M). Since accretion rates of thousands, or
even hundreds of \msun/yr are not sustainable, the latter scenario can occur
only if the SFR decreases significantly. In either cases, we predict a quick
decrease in the SFR of our sources. The presence of shocks and outflowing
gas and the moderate AGN-driven radio activity might be the expression of an
AGN-induced feedback that will cause the predicted SFR to decrease.
Such a mechanism is predicted in numerous current evolutionary
models~\citep{silk05,croton06b}, but it is more often invoked in AGNs at lower
redshifts and luminosities. Our results suggest that radio feedback might be
also important in high-$z$ QSOs~\citep[see also][]{sajina08,polletta08b}.

\section{Summary}

We analyze the multi-wavelength spectral energy distributions of two
obscured QSOs at high-$z$ discovered in the CFHTLS-D1/SWIRE survey, SWIRE4
J022550.67$-$042142.37 ($z$=3.867) and SWIRE4 J022513.92$-$043420.24
($z$=3.427). Both sources benefit from multi-wavelength data available
in the field and from mm data from MAMBO and optical and NIR spectroscopy
obtained as part of a dedicated follow up program.

Their large mm fluxes, $>$4\,mJy, imply IR luminosities
$\sim$10$^{12.5-13.3}$\,\lsun, and SFRs$\sim$500--3000\,\msun/yr. The two
sources also show powerful AGN activity, mostly seen in the MIR
(L(MIR)$_{AGN}\geq$10$^{13}$\,\lsun) and in the optical emission lines, and
probably also at radio wavelengths. Their MIR emission is mainly due to
AGN-heated hot dust and corresponds to a luminosity comparable to or higher
than the starburst luminosity. The MIR SEDs are redder and more luminous
than observed in other mm-detected obscured QSOs (see
\S~\ref{comparison} and Figure~\ref{multi_seds}). 

Their optical and NIR spectra are characterized by faint or absent
continuum emission and strong emission lines typical of AGNs. The optical
spectrum of SW022550 suggests either a high metalicity, $Z\sim4Z_{\odot}$,
or the presence of shocks. The comparison between the rest-frame ultraviolet
(UV) and optical spectra indicates the presence of outflowing material with a
velocity of $\sim$500\,\kms. The red and faint NIR emission compared to the
MIR emission indicates that the AGN is extincted by dust with \av$>$4. The
faint X ray emission, the large hardness ratio, and the small X ray/\oiii,
and X ray/MIR luminosity ratios suggest that their X ray emission might be
absorbed by a gas column density $\geq$10$^{24}$\,\cm2, and thus that they
are Compton-thick QSOs.

The sources show some hybrid properties in between those of type 1 and
type 2 AGNs, i.e. a broad component in some UV emission lines, blue UV
colors, and a significant emission from hot dust, likely at the sublimation
radius. We suggest that these properties are explained by the presence of
scattered light at UV wavelengths, and that a hot dust component might be
visible, perhaps associated with the NLR and with the scattering medium.

The scattering scenario will be tested with planned spectropolarimetric
observations on the VLT of SW022550. With those data we will investigate
whether the UV emission of SW022550 is dominated by a scattered component,
and measure the scattered fraction.

The composite (starburst and AGN) nature of our sources suggests
that they might be experiencing the transition phase between the starburst
and the QSO stage predicted by some evolutionary
models~\citep[e.g.][]{dimatteo05}. Since both sources already lie on the
local $M_{BH}-M_{bulge}$ relation, we predict a quick decrease in their SFR
in order to remain on such a relation as they evolve. Both
sources are characterized by AGN-driven moderate radio activity, providing
support to evolutionary models invoking radio feedback as star formation
quenching mechanism even in powerful QSOs~\citep{silk05,croton06a}.

\begin{acknowledgements}

We thank the referee, R. Antonucci, for a very careful reading and helpful
comments and suggestions. MP is grateful to S. Andreon for providing a list
of clusters around our sources. MP is also grateful to M. Bondi for
providing the VLA image of SW022513. MP thanks N. Fiolet for the final
validation of the MAMBO measurements. MP is particularly thankful to N.
Miller and C. Norman for providing the radio data for CDF-202, and for
helpful discussions. MP thanks S. H\"{o}nig for the permission to use the
torus models. CSS is grateful to S. Croom for providing the AUTOZ code. MP
acknowledges financial support from the Marie-Curie Fellowship grant
MEIF-CT-2007-042111. The AAOmega observations have been funded by the
Optical Infrared Coordination network (OPTICON), a major international
collaboration supported by the Research Infrastructures Programme of the
European Commission's Sixth Framework Programme. This work is based on
observations made with the {\it Spitzer Space Telescope}, which is operated
by the Jet Propulsion Laboratory, California Institute of Technology under
NASA contract 1407. We are thankful to the IRAM staff for their support with
the observations and the data reduction.

\end{acknowledgements}

\end{document}